\documentclass[pra,preprint,aps]{revtex4}

\usepackage{graphicx}

\begin{document}
\newcommand{\be}{\begin{equation}}
\newcommand{\ee}{\end{equation}}
\newcommand{\ben}{\begin{eqnarray}}
\newcommand{\een}{\end{eqnarray}}
\newcommand{\ra}{\rangle}
\newcommand{\la}{\langle}
\newcommand{\ov}{\overline}
\newcommand{\kn}{| n \rangle}
\newcommand{\bn}{ \langle n |}
\newcommand{\til}{\tilde}
\newcommand{\iii}{\'{\i}}

\newcommand{\pp}{p}
\newcommand{\kc}{| c \rangle}
\newcommand{\bc}{ \langle c |}
\newcommand{\kz}{| z \rangle}
\newcommand{\bz}{ \langle z|}
\newcommand{\kcz}{| \til{c}_z \rangle}
\newcommand{\bcz}{ \langle \til{c}_z |}

\newcommand{\lan}{  \lambda}
\newcommand{\kp}{| \pp \rangle}
\newcommand{\bp}{ \langle  \p2  |}
\newcommand{\ATA}{\hat{A}^\dagger \hat{A}}
\newcommand{\AAT}{  \hat{A} \hat{A}^\dagger}
\newcommand{\AT}{\hat{A}^\dagger}
\newcommand{\AO}{\hat{A}}
\newcommand{\sn}{\sum^M_{n=1}}
\newcommand{\snn}{\sum^N_{n=1}}
\newcommand{\RN}{{\cal {R}}^N}
\newcommand{\RM}{{\cal {R}}^M}
\newcommand{\kpn}{ | \psi_n \rangle}
\newcommand{\bpn}{\langle \psi_n|}
\newcommand{\kfn}{ | \phi_n \rangle}
\newcommand{\bfn}{\langle \phi_n|}
\newcommand{\kfo}{ | \ov{\phi}_n \rangle}
\newcommand{\bfo}{\langle \ov{ \phi}_n|}

\newcommand{\kf}{ | \ov{f} \rangle}
\newcommand{\kfe}{ | f^o \rangle}
\newcommand{\nn}{\nonumber \\ }
\newcommand{\nd}{\noindent}

\newcommand{\ton}{ \,;\, n=1,\ldots,N}
\newcommand{\toi}{ \,;\, i=1,\ldots,M}

\newcommand{\NC}{\hat{N}^\bot}
\newcommand{\NU}{Null(\AO)}
\newcommand{\NUC}{Null^\bot(\AO)}
\newcommand{\RA}{ Range(\AO)}
\newcommand{\PN}{\hat{P}_{N}}
\newcommand{\PNC}{\hat{P}_{N^{\bot}}}
\newcommand{\PR}{\hat{P}_{R}}

%\begin{document}
\draft
\title{On the Connection Between Entanglement
and the Speed of Quantum Evolution}
\author{J. Batle$^1$,  M. Casas$^1$,
 A. Plastino$^{2}$, and A. R. Plastino$^{3}$}

\affiliation{
$^1$Departament de F\iii sica, Universitat de les Illes Balears and IMEDEA-CSIC,
07122 Palma de Mallorca, Spain \\
$^2$National University La Plata-CONICET,
  C.C. 727, 1900 La Plata, Argentina\\
$^3$Department of Physics, University of Pretoria,
  0002 Pretoria, South Africa}

\date{\today}

%\maketitle

\begin{abstract}

\nd It has been recently pointed out [V. Giovanetti, S. Lloyd, and
L. Maccone, Europhys. Lett., {\bf 62} pp. 615-621 (2003)] that, for
certain classes of states, quantum entanglement enhances the
``speed" of evolution of composite quantum systems, as measured by
the time a given initial state requires to evolve to an orthogonal
state. We provide here a systematic study of this effect for pure
states of bipartite systems of low dimensionality, considering both
distinguishable (two-qubits) subsystems, and systems constituted of
two indistinguishable particles.

\noindent
 Pacs: 03.67.-a; 89.70.+c; 03.65.-w; 02.50.-r

\end{abstract}

\maketitle
%\vspace{.5cm}

%\vskip 5mm \noindent \hskip 2cm Keywords: Quantum Entanglement;
%classical entanglement; Quantum Information Theory
\newpage

\section{\bf Introduction}

Entanglement is one of the most fundamental 
features of the quantum description of Nature 
\cite{GLM03a,GLM03b,NC00,LPS98,BEZ00}. 
In recent years it has been the 
focus of intense research efforts  
\cite{GLM03a,GLM03b,NC00,LPS98,BEZ00,ABHHRWZ01,W98,TLB01,BCPP02b,BCPP02a,BPCP03,BCPP05,E02}. 
A state of a composite 
quantum system is called ``entangled" if it can not be 
represented as a mixture of factorizable pure states. 
Otherwise, the state is called separable.  Entanglement 
constitutes a physical resource that lies at 
the basis of important quantum information 
processes \cite{NC00,LPS98,BEZ00,ABHHRWZ01} 
such as quantum teleportation, superdense coding, 
and quantum computation.

 Entanglement is essential in connection both with 
 our basic understanding of quantum mechanics, 
 and with some of its most revolutionary (possible)
 technological applications. Consequently, it is 
 imperative to investigate in detail the relationships 
 between entanglement and other aspects of quantum 
 theory. In particular, it is of clear interest to 
 explore the role played by entanglement in the 
 dynamical evolution of composite quantum systems.
 It was recently discovered by Giovannetti, Lloyd, and
 Maccone \cite{GLM03a,GLM03b} that, in certain cases,
 entanglement helps to ``speed up" the time evolution
 of composite systems. The problem of the ``speed" 
 of quantum evolution has aroused considerable
 interest recently, because of its relevance
 in connection with the physical limits imposed
 by the basic laws of quantum mechanics on the
 speed of information processing and information
 transmission \cite{ML98,CD94,L00}.
 
 The aim of the present contribution 
 is to investigate in detail, for bipartite systems of low 
 dimensionality, the connection between entanglement and 
 the speed of quantum evolution. We are going to  focus 
 our attention on (i) two qubits (distinguishable) systems 
 and (ii) bosonic or  fermionic composite (bipartite) 
 systems of lowest dimensionality.

\section{Two entangled distinguishable particles}

 We are going to investigate first the case of two equal but 
 distinguishable subsystems evolving under a local Hamiltonian.
 Let us then consider a two qubits system whose evolution is governed
 by a (local) Hamiltonian 
 
 \be \label{haloloco}
 H \, = \, H_A \otimes I_B + I_A \otimes H_B,
 \ee

 \noindent
 where $H_{A,B}$  have eigenstates $|0\rangle$ and $|1\rangle$ with 
 eigenvalues $0$ and $\epsilon$, respectively. That is, the eigenstates of
 $H$ are $|00\rangle$, $|01\rangle$, $|10\rangle$, and $|11\rangle$,
 with eigenvalues respectively equal to $0$, $\epsilon$ (twofold degenerate) 
 and $2\epsilon$. For pure states $|\Psi \rangle  $ of our composite system
 the natural measure of entanglement is the usual reduced von Neumann 
 entropy $S[\rho_{A,B}] = -Tr_{A,B} (\rho_{A,B} \ln \rho_{A,B}) $ 
 (of either particle $A$ or particle $B$) where $\rho_{A,B} = Tr_{B,A} 
 (|\Psi \rangle \langle \Psi |)$. It is convenient for our present purposes
 to use, instead of $S(\rho_{A,B})$ itself, the closely related 
 ${\it concurrence \,\, C}$, given by
 
 \be \label{concurre1}
 C^2=4 \det \rho_{A,B}. 
 \ee

\noindent
Both the entanglement entropy $S[\rho_{A,B}]$ 
and the concurrence $C$ are preserved under 
the time evolution determined by the local 
Hamiltonian (\ref{haloloco}). Given an 
initial state

\begin{equation} \label{instate}
|\Psi(t=0)\rangle = c_0 |00\rangle+c_1 |01\rangle+c_2 |10\rangle+c_3 |11\rangle,
\end{equation}

\noindent
its concurrence is,

\be \label{concurre2}
C^2 \, = \, 4|c_0c_3-c_1c_2|^2.
\ee

\noindent
The overlap between the initial state (\ref{instate}) 
and the state at time $t$ is given by

\begin{equation} \label{overlape}
\langle\Psi(t)|\Psi(t=0)\rangle = |c_0|^2+(|c_1|^2+|c_2|^2)z+|c_3|^2z^2,
\end{equation}

\noindent 
where
$z \equiv {\rm exp}(i \epsilon t/\hbar) \equiv {\rm exp}(i\alpha)$, that is, 
$\alpha=\frac{t\epsilon}{\hbar}$.

\noindent
Thus, the condition for the state at time $t$ to be orthogonal to the
initial state is,

\be
P(z) \, = \, |c_0|^2+(|c_1|^2+|c_2|^2)z+|c_3|^2z^2 = 0.
\ee

\noindent
The above polynomial equation  can be cast as,

\be
|c_3|^2 (z-z_1) (z-z_2) = 0,
\ee

\noindent
where $z_1$ and $z_2$ are the roots of $P(z)$. If the
initial state $(\ref{instate})$ is to evolve to an 
orthogonal state, then the two roots of $P(z)$
have to be two (complex conjugate) numbers of modulus
equal to one. That is $z_{1,2} = \exp(\pm i \alpha)$. 
In that case we shall have,

\begin{eqnarray}
|c_0|^2=|c_3|^2&=&\Gamma, \cr
|c_1|^2+|c_2|^2&=&-2\Gamma {\rm cos}\alpha. 
\end{eqnarray}

\noindent
Appropriate normalization of the initial
state also implies that the concomitant 
coefficients can be parameterized as,

\begin{eqnarray}
|c_0|^2&=&|c_3|^2=\Gamma, \cr
|c_1|^2&=&-2\delta\Gamma {\rm cos}\alpha \cr
|c_2|^2&=&-2(1-\delta)\Gamma {\rm cos}\alpha, 
\end{eqnarray}

\noindent with $\Gamma=\frac{1}{2(1-{\rm cos}\alpha)}$ and 
$\alpha \in [\frac{\pi}{2},\frac{3\pi}{2}]$, $\delta \in [0,1]$. 
In other words, we have $\alpha={\rm arccos}(\frac{2\Gamma-1}{2\Gamma})$.

\noindent
The initial state's energy mean value and energy 
uncertainty are, respectively,  

\begin{eqnarray} \label{medenerg}
E \, &=& \, \langle H \rangle \, = \, 
\epsilon(|c_1|^2+|c_2|^2)+2\epsilon |c_3|^2=\epsilon \cr
\Delta E \, &=& \, \sqrt{\langle H^2 \rangle - \langle H \rangle^2} 
\, = \, \big(\epsilon^2(|c_1|^2+|c_2|^2)+4\epsilon^2 
|c_3|^2  - \epsilon^2
\big)^{\frac{1}{2}}=\epsilon \sqrt{2\Gamma}. 
\end{eqnarray}

\noindent 
The time $\tau$ required to evolve into an orthogonal state admits
the lower bound \cite{GLM03a,GLM03b},

\begin{equation} \label{Tmin} 
T_{min}=max\bigg(\frac{\pi\hbar}{2E},\frac{\pi\hbar}{2\Delta E}\bigg),
\end{equation}

\noindent
which, together with equations (\ref{medenerg}), lead to

\begin{equation} \label{Tminina} 
T_{min}=
\frac{\pi\hbar}{2\epsilon\sqrt{2\Gamma}}.
\end{equation}

\noindent
The concurrence of the (pure) state under 
consideration, defined as $C^2= 4|c_0c_3-c_1c_2|^2$ 
(see equations (\ref{concurre1}-\ref{concurre2})) is 

\be \label{concurre3}
C^2=4 \Big|\Gamma-e^{i\phi}
\sqrt{\delta(1-\delta)}
2\Gamma{\rm cos}\alpha \Big|^2.
\ee

\noindent
The modulus of the coefficients $c_i$ are completely determined
by the two parameters $\alpha$ (or $\Gamma$) and $\delta$. 
The dependence of $C^2$ on the phases of the coefficients 
$c_i$ can be absorbed into one single phase $e^{i\phi}$, 
thus incorporating a new parameter $\phi$ into the expression 
(\ref{concurre3}) for $C^2$. 

After some algebra, the expressions for the minimum and 
maximum values for the evolution time $\tau$ that 
are actually realized for states of a given concurrence 
$C^2$, read

\begin{equation} \label{tautau}
\frac{\tau}{T_{min}(\Gamma)}=\frac{2}{\pi}\sqrt{2\Gamma}\,{\rm arccos}
\left(\frac{2\Gamma-1}
{2\Gamma} \right),
\end{equation}
  
\noindent
where the maximum evolution time for a fixed $C^2$ 
(or a fixed $C$) corresponds to $\Gamma=\sqrt{C^2}/2$, 
while the minimum one to $\Gamma=(1+\sqrt{C^2})/4$. 
The two curves in the $(C,\tau/T_{min})$-plane 
corresponding, for each value of $C$, to the states
with maximum and minimum $\tau/T_{min}$ are depicted
in Figure 1. All states that eventually evolve into 
an orthogonal state (that is, states characterized 
by different $\delta$'s and $\phi$'s) lie between 
these two curves. Some important features of the 
connection between entanglement and speed of 
evolution (for two qubits) transpire from 
Figure 1. First, we see that both the 
maximum and the minimum times required to
reach an orthogonal state are monotonously
decreasing functions of the concurrence. 
Second, the difference between these 
maximum and the minimum evolution times
(that is, the range of possible values 
for the time required to evolve 
to an orthogonal state) also 
decreases with increasing concurrence. 
Third, the lower bound for the evolution
time to an orthogonal state is saturated 
by (and only by) the maximally entangled
states ($C=1$). These features provide 
further support to the idea that 
entanglement tends to ``speed up" 
quantum evolution.

\section{Two entangled indistinguishable particles}
 
 Here we are going to explore the connection between 
 entanglement and the speed of quantum evolution for 
 systems constituted by two indistinguishable particles.
 In this case the concept of entanglement exhibits some
 extra subtleties, as compared with the case of 
 distinguishable subsystems. When dealing with 
 indistinguishable particles, the correlations
 that arise purely from the concomitant
 statistics (either fermionic or bosonic) 
 do not constitute a useful resource and,
 consequently, must not be regarded as 
 contributing to the amount of entanglement 
 of the system's state \cite{E02}.
 A useful formalism to describe the entanglement of
 systems consisting of identical particles, that takes
 into account the above remarks, has been 
 advanced by  Eckert {\it et al.} in \cite{E02}. For 
 two identical bosons, the system of lowest 
 dimensionality exhibiting the phenomenon of entanglement 
 is a pair of bosons with a two dimensional single 
 particle Hilbert space. The simplest fermionic 
 system endowed with entanglement is a system 
 of two fermions with a three dimensional
 single particle Hilbert space.

\subsection{Bosons}

Using the second quantization formalism, 
the general (pure) state of two bosons 
(with a two-dimensional single particle 
Hilbert space) can be written under the 
guise \cite{E02}

\begin{equation} \label{sboson}
|V\rangle = \sum_{i,j=0}^{1} v_{ij} b_i^{\dagger}b_j^{\dagger} |0\rangle,
\end{equation}

\noindent  
where $b_i^{\dagger}$ and $b_i$ denote
bosonic creation and anhilation operators, 
the coefficients $v_{ij}$ constitute 
the symmetric matrix

\begin{equation}
\hat V= \left(
\begin{array}{cccc}
v_{00} & v_{01} \\
v_{10} & v_{11} \end{array} \right).
\end{equation}

\noindent That is, $v_{ij}=v_{ji}$. Normalization 
imposes the condition 
$2\sum_{i,j=0}^{1} |v_{ij}|^2=1$. 

The Hamiltonian associated with two 
non-interacting bosons is,

\begin{equation} \label{hamil}
\hat H = \sum_{k=0}^{1} \epsilon_k b_k b_k^{\dagger},
\end{equation}

\noindent where $b_0^{\dagger}|0\rangle$ is the single particle 
ground state with energy $\epsilon_0 = 0$, and  
$b_1^{\dagger}|0\rangle$ is the single particle excited 
state with energy $\epsilon_1 = \epsilon$. The state 
(\ref{sboson}) evolves according to the time-dependent 
Schroedinger equation, 

\begin{equation}
i\hbar \frac{d}{dt} |V(t)\rangle=\hat H |V(t)\rangle = 
\sum_{i,j=0}^{1} (\epsilon_i+\epsilon_j)
v_{ij}(t) b_i^{\dagger}b_j^{\dagger} 
|0\rangle,
\end{equation}

\noindent The general solution of this evolution equation
is given by the time dependent coefficients,

\begin{equation}
v_{ij}(t) = v_{ij}(0)\,e^{-i\frac{(\epsilon_i+\epsilon_j)}{\hbar}t}.
\end{equation}

\noindent The time $\tau $ required to evolve into an 
orthonormal state is

\begin{equation} \label{orthoboson}
\langle V(0)|V(\tau)\rangle=2 \sum_{i,j=0}^{1} |v_{ij}(0)|^2\,
e^{-i\frac{(\epsilon_i+\epsilon_j)}{\hbar}\tau}=0.
\end{equation}

\noindent Setting $z \equiv e^{-i\frac{\epsilon \tau}{\hbar}}
=e^{-i\alpha}$, the orthogonality condition 
(\ref{orthoboson}) can be recast as a 
polynomial equation in $z$, that has 
to admit roots of modulus equal to $1$.
From this last requirement, and taking into 
account the symmetries in the coefficients 
$v_{ij}$, it follows that the coefficients
can be parameterized as,

\begin{eqnarray}
|v_{00}|^2&=&\Gamma \cr
|v_{01}|^2&=&-\Gamma {\rm cos}\alpha \cr
|v_{11}|^2&=&\Gamma,
\end{eqnarray}

\noindent 
with $\Gamma > 0$ and $\alpha \in [\pi/2, 3\pi/2] $.
The normalization constraints also implies that 
$\Gamma=\frac{1}{4(1-{\rm cos}\alpha)}$. The 
expectation values of the energy and its square
read

\begin{eqnarray}
E&=&\langle H \rangle = 
2\sum_{i,j=0}^{1} |v_{ij}(0)|^2\,
(\epsilon_i+\epsilon_j) = \epsilon \cr
\langle H^2 \rangle&=& 
2\sum_{i,j=0}^{1} |v_{ij}(0)|^2\,
(\epsilon_i+\epsilon_j)^2=
(4 \Gamma + 1)\epsilon^2,
\end{eqnarray}

\noindent and consequently the minimum 
evolution time (\ref{Tmin}) is 

\be
T_{min}=\frac{\pi \hbar}{2\epsilon}.
\ee

\noindent
The formula for the concurrence in 
the two-boson case is \cite{E02}

\be
C_{B}=4|v_{00}v_{11}-v_{01}^2|, 
\ee

\noindent
which is clearly time-independent. 

For a given value of the concurrence, the minimum 
and maximum times for evolution to an orthogonal 
state can be obtained in the same way as in the
case of two distinguishable qubits. The equations
relating the minimum and maximum times with the 
concurrence are, respectively, 

\be \label{temin}
C=\frac{1+{\rm cos}\left( \alpha_{\rm min} \right)}
{1-{\rm cos}\left(\alpha_{\rm min}\right)}
\ee

\noindent
and

\be \label{temax}
C=\frac{1}{1-{\rm cos}\left(\alpha_{\rm max}\right)},
\ee

\noindent
where $\alpha_{min,max} = 
\exp(-i \epsilon \tau_{min,max} / \hbar)$.
The curves associated with equations 
(\ref{temin}-\ref{temax}), depicting the 
extremum evolution times as a function of 
$C$, are exhibited in Figure 2.

Comparing Figure 2 with Figure 1 we  see 
that the same general trends exhibited by a
system of two distinguishable qubits are also 
observed in the case of two identical boson.

\subsection{Fermions}

Now we are going to study a system of two identical 
fermions with a three dimensional single particle
Hilbert space. In second quantization notation,
the general (pure) state of such a system is,

\begin{equation} \label{sfermion}
|W\rangle = \sum_{i,j=0}^{3} w_{ij} f_i^{\dagger}f_j^{\dagger} |0\rangle,
\end{equation}

\noindent where $f_i^{\dagger}$ and $f_i$ denote
fermionic creation and anhilation operators, 
respectively, and the coefficients $w_{ij}$ 
constitute the anti-symmetric matrix

\begin{equation}
\hat W= \left(
\begin{array}{cccc}
0 & w_{01} & w_{02} & w_{03} \\
w_{10} & 0 & w_{12} & w_{13} \\
w_{20} & w_{21} & 0 & w_{23} \\
w_{30} & w_{31} & w_{32} & 0 \end{array} \right).
\end{equation}

\noindent That is, $w_{ij}=-w_{ji}$. 
Normalization imposes the condition 
$\sum_{i,j=0}^{3} |w_{ij}|^2=1/2$. 
The Hamiltonian describing two non-interacting 
particles is given by,

\begin{equation} \label{hamilhamil}
\hat H = \sum_{k=0}^{1} \epsilon_k f_k f_k^{\dagger},
\end{equation}

\noindent
and the coefficients

\begin{equation}
w_{ij}(t) = w_{ij}(0)\,
e^{-i\frac{(\epsilon_i+\epsilon_j)}{\hbar}t},
\end{equation}

\noindent 
describe a general solution of the concomitant time depending
Schroedinger equation. Let $z \equiv 
e^{-i\frac{\epsilon \tau}{\hbar}}=
e^{-i\alpha}$. The evolution time to an orthogonal state 
follows from the condition 

\begin{eqnarray} \label{ferpolizet}
\langle W(0)|W(\tau)\rangle &=& 2 \sum_{i,j=0}^{3} |w_{ij}(0)|^2\,
e^{-i\frac{(\epsilon_i+\epsilon_j)}{\hbar} \tau} \nn
&=& 4 z \left( |w_{01}|^2+|w_{02}|^2\,z+
(w_{03}|^2+|w_{12}|^2)\,z^2+
|w_{13}|^2\,z^3+|w_{23}|^2\,z^4\right) \nn
\, &=& \, 0.  
\end{eqnarray}

\noindent The polynomial equation (\ref{ferpolizet}) 
may have either (i) fourth real roots, (ii) two real roots 
and two complex (complex conjugated) roots, or 
(iii) two pairs of complex conjugated roots. 
Since we are interested in solutions of the type 
$e^{-i\frac{\epsilon}{\hbar}\tau}$, the most general 
case of interest is (iii). Consequently, the 
two solutions of (\ref{ferpolizet}) corresponding 
to (positive) times of evolution into an orthogonal 
state are of the form $z_1 \equiv e^{-i\alpha}$ and 
$z_2 \equiv e^{-i\beta}$. Taking into account the 
antisymmetric nature of $w_{ij}$, we get 
the following relations

\begin{eqnarray}
|w_{01}|^2 &=& x \\
|w_{02}|^2 &=& -2\,x({\rm{cos}}\alpha + {\rm{cos}}\beta)\\
|w_{03}|^2+|w_{12}|^2 &=& 2\,x(1+2{\rm{cos}}\alpha\,{\rm{cos}}\beta)\\
|w_{13}|^2 &=& -2\,x({\rm{cos}}\alpha + {\rm{cos}}\beta)\\
|w_{23}|^2 &=& x,
\end{eqnarray}

\noindent where the value of the parameter $x$ 
is determined by the normalization requirement. 
We want to find the fastest solution to the 
first orthogonal state. The time $\tau $ 
required to reach an orthogonal state is 

\be
\tau= \frac{\hbar}{\epsilon}  
\times min(\alpha,\beta).
\ee

Let us consider the case $\beta=\pi$. Then, the
time required to arrive to an orthogonal state
is equal to $\tau = \hbar \alpha/\epsilon$
and the coefficients characterizing the quantum
state are,

\begin{eqnarray}
|w_{01}|^2 &=& \frac{1}{32\,(1-{\rm{cos}}\alpha)} \\
|w_{02}|^2 &=& \frac{1}{16} \\
|w_{03}|^2+|w_{12}|^2 &=& \frac{1-2{\rm{cos}}\alpha}{16\,(1-{\rm{cos}}\alpha)} 
\\
|w_{13}|^2 &=& \frac{1}{16} \\
|w_{23}|^2 &=& \frac{1}{32\,(1-{\rm{cos}}\alpha)}, 
\end{eqnarray}

\noindent 
with the obvious condition $ \cos \alpha < 1/2$
(that is, $\alpha \in [\pi/3,\pi ]$). 
The energy and energy square expectation values 
read

\begin{eqnarray}
E&=& \langle H \rangle = 
2\sum_{i,j=0}^{3} 
|w_{ij}(0)|^2\,(\epsilon_i+\epsilon_j) = 
3 \epsilon \cr
\langle H^2 \rangle &=&
2\sum_{i,j=0}^{3} |w_{ij}(0)|^2\,
(\epsilon_i+\epsilon_j)^2 = \, \frac{\epsilon^2}{2} 
\left[\frac{21-19\cos \alpha}{1 - \cos \alpha} \right],
\end{eqnarray}

\noindent and, consequently, the minimum 
evolution time (\ref{Tmin}), after 
some calculation, is

\be
T_{min} \, = \, \frac{\pi \hbar}{2 \epsilon} \,
\sqrt{\frac{2(1 - \cos \alpha)}{3 - \cos \alpha}}.
\ee 

\noindent
%Notice that this formula 
%also holds for arbitrary values of $\alpha$ 
%and $\beta$.
The formula for the concurrence in the 
two-fermion case is \cite{E02}

\be \label{fermiconc}
C_{F}=8|w_{01}w_{23}-w_{02}w_{13}+w_{03}w_{12}|,
\ee

\noindent
which is clearly time-independent (for the Hamiltonian
(\ref{hamilhamil})).

One can check that the lowest value of 
$\tau/T_{min}$ corresponds to 
$\cos \alpha = 1/2$. That is, the 
state closest to saturate the 
lower bound for the time required
to reach an orthogonal state is
given by $\alpha=\pi/3$. In this
case the fermionic concurrence 
reads 

\begin{equation} \label{C_F}
C_F = \frac{|e^{i\phi_{01}+i\phi_{23}}-e^{i\phi_{02}+i\phi_{13}}|}{2}, 
\end{equation}

\noindent where $\phi_{ij}$ denotes the phase
associated with the coefficient $w_{ij}$.
Now, with an appropiate choice of the $\phi's$, 
we can make (\ref{C_F}) either $0$ or $1$. 
In other words, among those states that saturate
the lower bound on the time to evolve to an orthogonal
state, there are states of zero entanglement, 
as well as maximum entangled states.

Figure 3 exhibits a plot in the $(C,\tau/T_{min})$-plane 
of a set of randomly generated states of two fermions 
that evolve to an orthogonal state. Each point represents 
one of those states. It transpires from the figure that 
for each value of the concurrence $C$ the time  
$\tau/T_{min}$ needed to reach an orthogonal state may 
adopt any value, from $\frac{1}{3} \sqrt{10}$ 
up to a maximum equal to $2$. 

We see that, as far as the connection between 
entanglement and the ``speed " of quantum evolution
is concerned, the behaviour of fermionic systems 
differs considerably  from the behavior of
systems consisting either of bosons, or of
distinguishable particles.

\section{Conclusions}

 We have explored, for bipartite systems of low 
 dimensionality, some aspects of the connection 
 between entanglement and the speed of quantum 
 evolution. We considered (i) two qubits (distinguishable) 
 systems  and (ii) systems composed of two (bosonic or 
 fermionic) identical particles, with single
 particle Hilbert spaces of lowest dimensionality.

 Our present results corroborate that 
 there is a clear correlation between the
 amount of entanglement and the speed of 
 quantum evolution for systems of two-qubits 
 and systems of two identical bosons. On 
 the contrary, such a clear correlation is 
 lacking in the case of systems of 
 identical fermions.

\vskip 3mm
{\bf Acknowledgements}
\vskip 3mm
This work was partially supported by the MEC grant BFM2002-03241 (Spain)
and FEDER (EU), by the Government of Balearic Islands and by CONICET
(Argentine Agency).

\newpage

\begin{figure}[t]
\begin{center}
\includegraphics*[scale=0.35,angle=270]{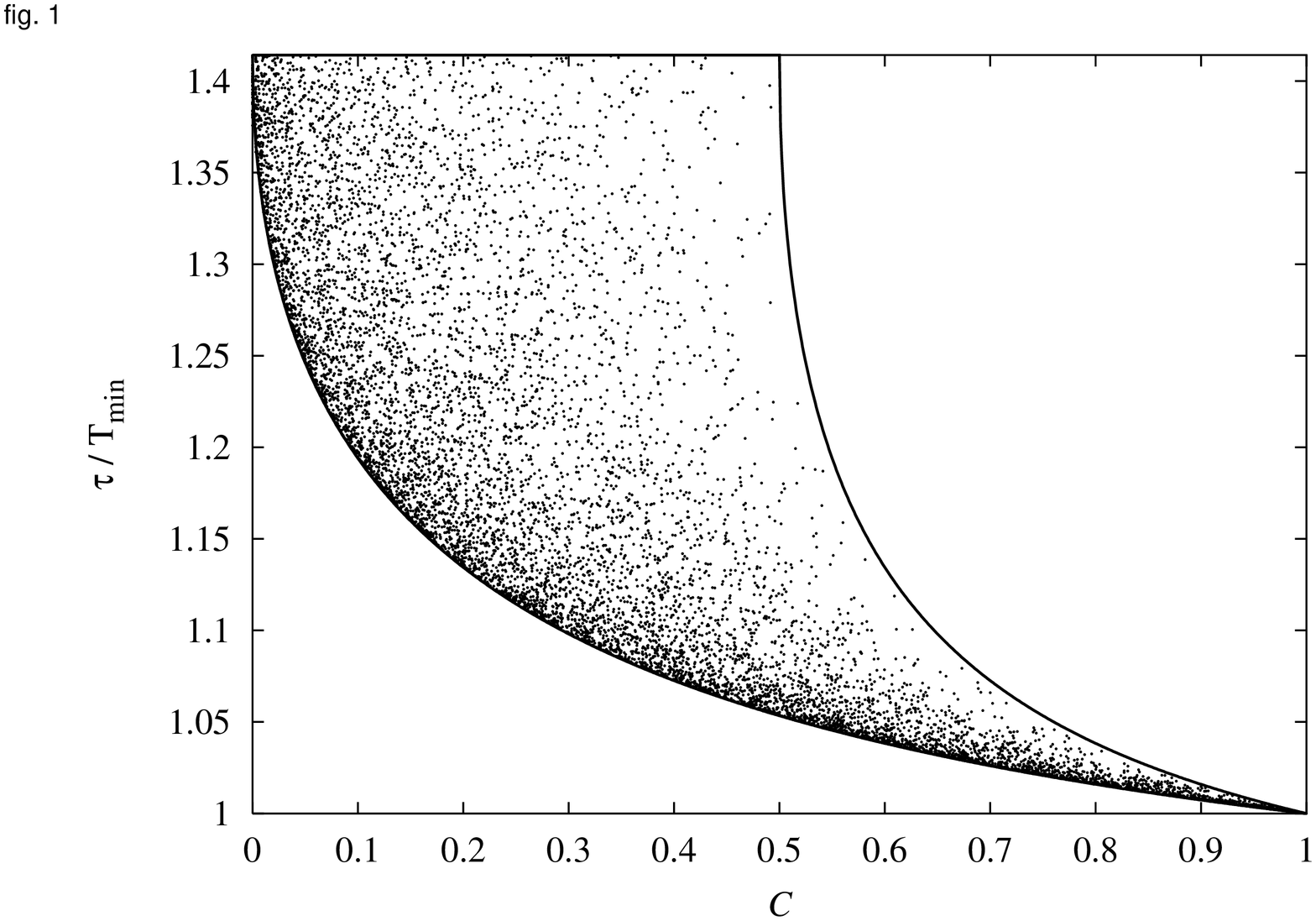}
\caption{
Curves in the $(C,\tau/T_{min})$-plane corresponding, 
for each value of $C$, to the states of
two (distinguishable) qubits with maximum and 
minimun $\tau/T_{min}$. The points represent 
randomly generated individual states that evolve 
to an orthogonal state. All depicted quantities
are dimensionless}
\end{center}
\label{fig1}
\end{figure}

\newpage

\begin{figure}[t]
\begin{center}
\includegraphics*[scale=0.35,angle=270]{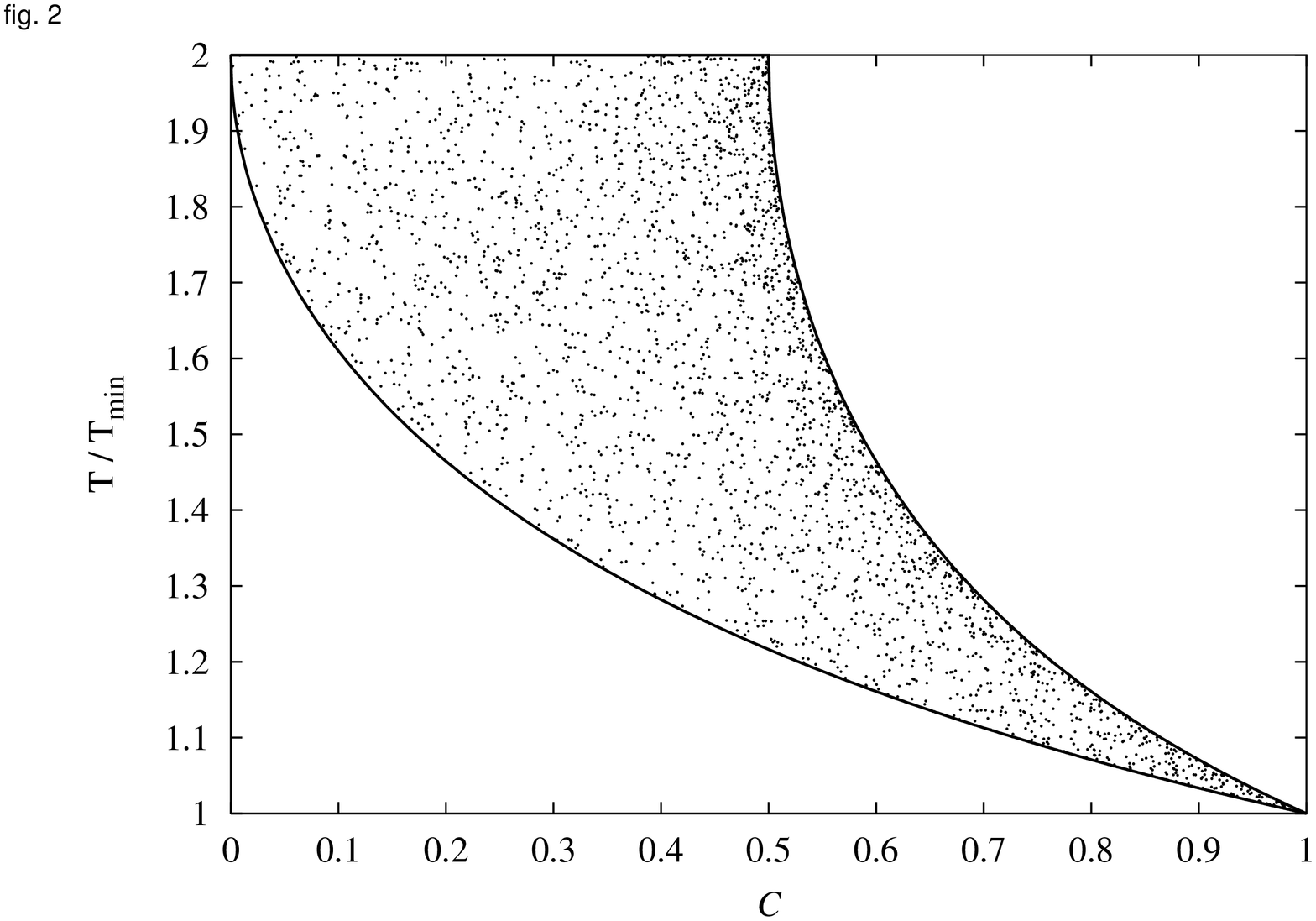}
\caption{Curves in the $(C,\tau/T_{min})$-plane 
corresponding, for each value of $C$, to the states 
of two bosons with maximum and minimun $\tau/T_{min}$.
The points represent randomly generated individual 
states that evolve to an orthogonal state.
All depicted quantities
are dimensionless}
\end{center}
\label{fig2}
\end{figure}

\newpage

\begin{figure}[t]
\begin{center}
\includegraphics*[scale=0.35,angle=270]{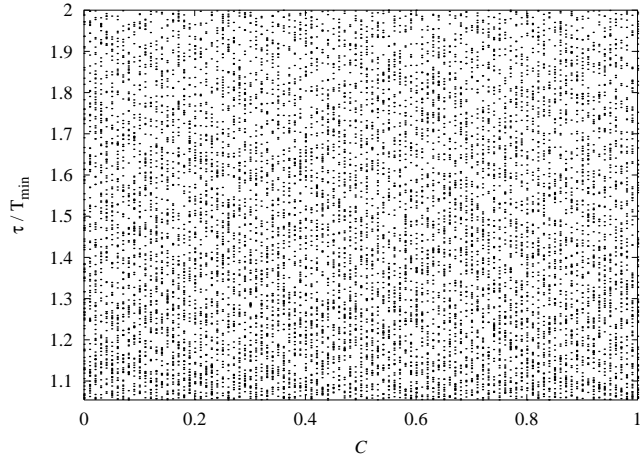}
\caption{Randomly generated states of two fermions that
evolve to an orthogonal state. Each point corresponds 
to one of those states, represented in the 
$(C,\tau/T_{min})$-plane. It transpires from the figure 
that for each value of the concurrence $C$ the time  
$\tau/T_{min}$ needed to reach an orthogonal state 
may adopt any value, from $\frac{1}{3}\sqrt{10}$
up to a maximum equal to $2$. All depicted quantities
are dimensionless}
\end{center}
\label{fig3}
\end{figure}

\end{document}